\documentclass[titlepage,preprint,nofootinbib,prd,nopacs,
tightenlines,byrevtex,amsmath]{revtex4}

\usepackage{graphicx}
\usepackage{bm}
\newcommand{\beq}{\begin{equation}}
\newcommand{\eeq}{\end{equation}}
\newcommand{\eq}[1]{Eq.~(\ref{#1})}
\newcommand{\crossed}[1]{#1\hspace{-.5em}\slash}

\begin{document}

\preprint{UK/09-07}
\title {Radiative-Recoil Corrections to Hyperfine Splitting:\\ Polarization Insertions in the Electron Factor}
\author {Michael I. Eides}
\altaffiliation{Also at Petersburg Nuclear Physics Institute,
Gatchina, St.Petersburg 188300, Russia}
\email{eides@pa.uky.edu, eides@thd.pnpi.spb.ru}
\affiliation{Department of Physics and Astronomy,
University of Kentucky,
Lexington, KY 40506, USA,}

\author{Valery A. Shelyuto}
\email{shelyuto@vniim.ru}
\affiliation{D. I.  Mendeleev Institute of Metrology,
St.Petersburg 190005, Russia}

\begin{abstract}
We consider three-loop radiative-recoil corrections to hyperfine splitting in muonium due to insertions of one-loop polarization operator in the electron factor. The contribution generated by electron polarization insertions is a cubic polynomial in the large logarithm of the electron-muon mass ratio. The leading logarithm cubed and logarithm squared terms are well known for some time. We calculated all single-logarithmic and nonlogarithmic radiative-recoil corrections of order $\alpha^3(m/M)E_F$ generated by the diagrams with electron and muon polarization insertions.

\end{abstract}


\maketitle

\section{Introduction}

\begin{figure}[tbh]
\center\includegraphics{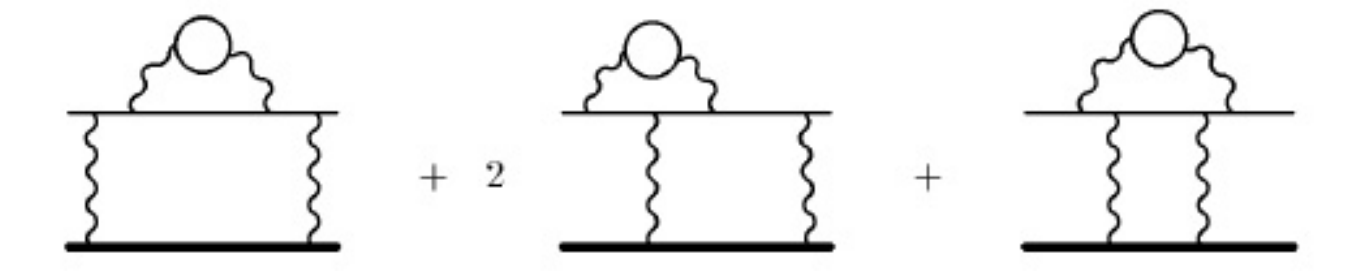}
\caption{\label{elline} Electron polarization insertions}
\end{figure}

\begin{figure}[tbh]
\center\includegraphics{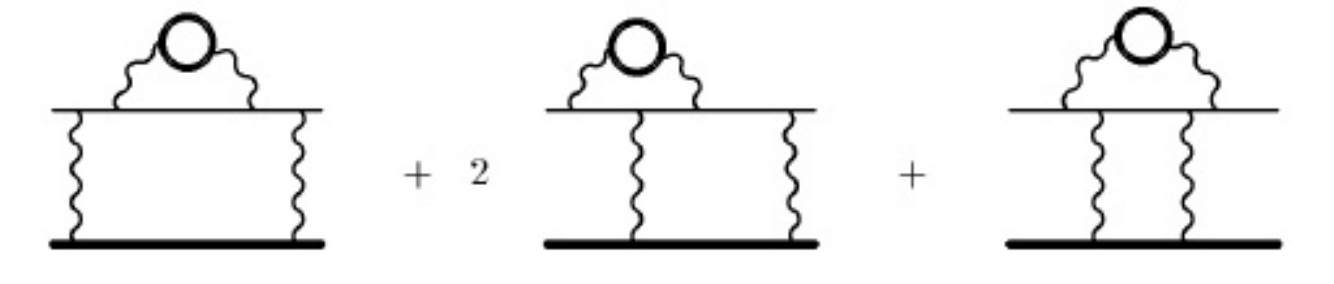}
\caption{\label{muline} Muon polarization insertions}
\end{figure}

Leading three-loop logarithm cubed and logarithm squared
radiative-recoil contributions  to hyperfine splitting (HFS) in
muonium were calculated long time ago (see, e.g., reviews in
\cite{egsbook,preprts}). Recently we started calculation of all
single-logarithmic and nonlogarithmic radiative-recoil correction
(see review in \cite{egscan2005}). Below we consider
single-logarithmic and nonlogarithmic radiative-recoil corrections
due to insertions of electron and muon polarization operators in the
radiative photon in Figs.~\ref{elline}, \ref{muline}.

\begin{figure}[tbh]
\center\includegraphics[height=3.cm]{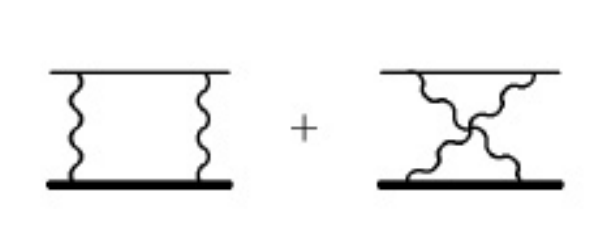}
\caption{\label{twoph} Two-photon exchanges}
\end{figure}

Three-loop diagrams in Figs.~\ref{elline}, \ref{muline} can be obtained from the diagrams with two-photon exchanges by insertion of radiative corrections in Fig.~\ref{twoph}. The two-photon diagrams in Fig.~\ref{twoph} produce the leading radiative-recoil correction when the loop momentum is much larger than the electron mass, and insertion of radiative correction can only increase the integration momentum. Therefore, calculating the diagrams in Figs.~\ref{elline}, \ref{muline} we may forget about external virtualities and calculate matrix elements in the scattering regime between the free electron and muon spinors. To turn the matrix element into contribution to HFS we multiply the scattering matrix element by the Coulomb-Schr\"odinger bound state wave function at the origin squared, and calculate difference between spin one and spin zero states. We use the Feynman gauge to obtain matrix elements of the gauge invariant sets of diagrams in Figs.~\ref{elline}, \ref{muline}. Each of the diagrams in Figs.~\ref{elline}, \ref{muline} contains polarization operator insertion in one of the radiative photon lines. We account for this insertion using the massive photon propagator for radiative photons (but not for exchanged photons) with the photon mass squared $\lambda^2=4m^2/(1-v^2)$ or $\lambda^2=4M^2/(1-v^2)$, where  $m$ and $M$ are the electron and muon masses, respectively. Insertion of the polarization operator in the radiative photon line is accompanied by an additional integration over velocity $v$ with the weight

\beq \label{velintweight}
\int_0^1 \frac{{dv}}{1-v^2}~v^2\biggl(1-\frac{v^2}{3}\biggr).
\eeq

Besides single-logarithmic and nonlogarithmic contributions the
diagrams Fig.~\ref{elline} generate also well known much larger nonrecoil and logarithm squared recoil contributions \cite{egsbook}. We will calculate the contributions of the diagrams Figs.~\ref{elline}, \ref{muline} with linear accuracy in the small electron-muon mass ratio $m/M$. In particular we will reproduce nonrecoil and logarithm squared recoil contributions what will serve as an additional check of our new results. The paper is organized as follows. In Section II we describe calculations of the diagrams with the electron polarizations in Fig.~\ref{elline}, and Section III  deals with the diagrams with the muon polarizations in Fig.~\ref{muline}. All results are collected in the last section.

\section{Electron Polarization Operator}

\subsection{Calculation of the Mass Operator Contribution}

We calculate matrix elements of each of the diagrams in Fig.~\ref{elline} separately. Respective integrals can be obtained by inserting radiative corrections in the expression for the contribution of the skeleton diagrams in Fig.~\ref{twoph} to HFS. Contribution of the diagram with the self-energy insertion in Fig.~\ref{elline} has the form (compare \cite{yaf1988})

\beq \label{elsigmatot}
\Delta \epsilon_{\Sigma}=\frac{3i}{8\pi^2\mu^2}\int_0^1 {dx}
\int_0^x {dy} \int_0^1
\frac{{dv}}{1-v^2}~v^2\biggl(1-\frac{v^2}{3}\biggr) \int
{\frac{d^4k}{k^4}}
\eeq
\[
\times\frac{2k^2}{k^4-\mu^{-2}k_0^2}
\frac{1}{-k^2+2k_0+a_1^2(x,y)-i0}\biggl[h_1(x,y)  k_0 -
\frac{h_2(x,y)}{3}(2k^2+k_0^2)\biggr]\equiv \Delta
\epsilon_{\Sigma1}+\Delta \epsilon_{\Sigma2},
\]

\noindent
where dimensionless energy $\Delta \epsilon_{\Sigma}$ is connected
with the energy shift $\Delta E_{\Sigma}$ by the relationship $\Delta
E_{\Sigma}=(\alpha^2(Z\alpha)/\pi^3)(m/M)E_F\Delta \epsilon_{\Sigma}$\footnote{The Fermi energy is defined as $E_F=(8/3)(Z\alpha)^4(m/M)m$.}, $\mu=m/(2M)$, $k$ is the dimensionless Minkowski exchanged momentum, $\Delta\epsilon_{\Sigma1}$ and $\Delta \epsilon_{\Sigma2}$ are the integrals corresponding to two terms in the square brackets, and

\beq \label{auxsigmele}
h_1(x, y)=\frac{1+x}{y},\quad
h_2(x, y)=\frac{1-x}{y}
\biggl[1-\frac{2(1+x)y}{x^2+\lambda^2 (1-x)}\biggr],\quad
a_1^2(x, y)=\frac{x^2+\lambda^2 (1-x)}{(1-x)y}.
\eeq

\noindent
The integration over $v$ in \eq{elsigmatot} takes care the electron loop that is included in the integrand via the finite mass $\lambda=\sqrt{4/(1-v^2)}$ of the radiative photon. We are going to calculate two leading terms in the expansion of the dimensionless energy $\Delta \epsilon_{\Sigma}$  in \eq{elsigmatot} over the small parameter $\mu$ (the term of order $1/\mu$ and the term independent of $\mu$).

\subsubsection{Nonrelativistic Contribution}

As a first step of calculations we obtain the leading term of order $1/\mu$
from the expression for $\Delta \epsilon_{\Sigma}$ in \eq{elsigmatot}. This term gives the leading nonrecoil contribution to HFS (recall
$\Delta E_{\Sigma}=(\alpha^2(Z\alpha)/\pi^3)2\mu\Delta
\epsilon_{\Sigma}$) and arises, as all nonrecoil contributions,
in the external field approximation. To extract the nonrecoil
contribution from the expression in \eq{elsigmatot} we take the
residue at the muon pole what with linear accuracy in $\mu$ reduces
to substitution

\beq
\frac{2k^2}{k^4-\mu^{-2}k_0^2}\to -2\pi i\mu\delta(k_0-\mu {\bm
k}^2).
\eeq

\noindent
Then we obtain the nonrelativistic for the muon contribution in the form

\beq \label{nonrelsigma}
\Delta \epsilon_{\Sigma}(NR)=\frac{3}{4\pi\mu}\int_0^1 {dx}
\int_0^x {dy} \int_0^1
\frac{{dv}}{1-v^2}v^2\biggl(1-\frac{v^2}{3}\biggr)
\eeq
\[
\times\int
\frac{{d^3k}}{{\bm k}^2[{\bm k}^2(1+2\mu)+a_1^2(x,y)]}\biggl[\mu h_1(x,y)
+ \frac23 h_2(x,y) \biggr].
\]

\noindent
This last integral contains both recoil and nonrecoil contributions.
The recoil contribution will be treated on equal  grounds with
other recoil contributions considered below. Integrating in
\eq{nonrelsigma} over momentum and expanding the result over the
small parameter $\mu$ we obtain

\beq \label{nonrelsigmares}
\Delta \epsilon_{\Sigma}(NR)= \int_0^1 {dx} \int_0^x {dy}
\int_0^1 \frac{{dv}}{1-v^2}v^2\biggl(1-\frac{v^2}{3}\biggr)
\biggl[\frac{1}{\mu}\frac{\pi}{a_1(x,y)}h_2(x,y)
\eeq
\[
+ \frac{\pi}{2a_1(x,y)} (3 h_1(x,y) - 2 h_2(x,y)) \biggr]
=\frac{0.4462}{\mu}   + 2.2372
\equiv \Delta
\epsilon_{\Sigma}^{(\mu)}(NR)+\Delta \epsilon_{\Sigma}^{(c)}(NR).
\]

\subsubsection{$\mu$ - and $c$ - Integrals}

We return now to calculation of the first two terms in the expansion
of the contributions to HFS in \eq{elsigmatot} over $\mu$. An attempt
to calculate the integral in \eq{elsigmatot} with the help of the
Feynman parameters leads to the integrands that do not admit
expansion over small parameter $\mu$ before integration. Therefore we
use another approach to calculation of the integral in
\eq{elsigmatot} (as well as to calculation of other integrals of this type below), and directly integrate over the exchanged momentum $k$ in
four-dimensional polar coordinates. After the Wick rotation and
integration over angles (and omitting some higher order terms in
$\mu$) we obtain integral representation for the contributions $\Delta
\epsilon_{\Sigma1}$ and $\Delta \epsilon_{\Sigma2}$ defined in
\eq{elsigmatot}. The integral representation for  $\Delta
\epsilon_{\Sigma1}$ (compare \cite{jetp92}) has the form

\beq \label{sigma1tot}
\Delta \epsilon_{\Sigma 1}=-3\int_0^1 {dx} \int_0^x
{dy} \int_0^1 \frac{{dv}}{1-v^2}v^2\biggl(1-\frac{v^2}{3}\biggr)
h_1(x,y)\int_0^{\infty}{dk^2}
\biggl\{\frac{1}{(k^2+a_1^2)^2}
\eeq
\[
\times\biggl(\mu k \sqrt{1+\mu^2k^2} - \mu^2
k^2 \biggr)
-\frac{1}{4k^2}\biggl[\frac{1}{k^2+a_1^2}\sqrt{(k^2+a_1^2)^2+4k^2}
-1\biggr]\biggr\},
\]

\noindent
and the integral representation for $\Delta \epsilon_{\Sigma2}$
(compare \cite{yaf1988}) has the form

\beq \label{sigma2tot}
\Delta \epsilon_{\Sigma 2}= \int_0^1 {dx} \int_0^x {dy} \int_0^1
\frac{{dv}}{1-v^2}v^2\biggl(1-\frac{v^2}{3}\biggr) h_2(x,y)
\int_0^{\infty}{dk^2}\biggl\{\frac{1}{k^2+a_1^2}
\eeq
\[
\times\biggl[\frac{1}{\mu k}\biggl(\sqrt{1+\mu^2k^2} \biggr)
- \frac{1}{2}\biggl(\mu k \sqrt{1+\mu^2k^2}- \mu^2 k^2
\biggr)\biggl]
\]
\[
+\biggl[-\frac{\sqrt{(k^2+a_1^2)^2+4k^2}}{(k^2+a_1^2)^2}
+\frac{1}{8k^2}\biggl(\sqrt{(k^2+a_1^2)^2+4k^2}-(k^2+a_1^2)
\biggr)\biggl]\biggr\}.
\]

\noindent
Each integrand in \eq{sigma1tot} and \eq{sigma2tot} is a sum of
$\mu$-dependent and $\mu$-independent integrals, and we would like to
calculate them separately. While both integrals are convergent, naive separation of $\mu$-dependent and $\mu$-independent terms in the integrands sometimes leads to  integrals that are ultravioletly divergent at large integration momenta $k$. For example, integral over $k$ of $\mu$-dependent term in the
integrand in \eq{sigma1tot} converges at large integration
momenta, while respective integral in \eq{sigma2tot} diverges. This
divergence arises because $\mu$-dependent terms in the integrand go
to $\mu$-independent constants at high momenta. In such cases we
redefine the $\mu$-dependent terms in the integrand by subtracting
the leading asymptotic constant,  and add this constant to the
$\mu$-independent terms in the integrand. A universal recipe for
such restructuring of the integrands in \eq{sigma1tot}, \eq{sigma2tot} is
described by the substitutions

\beq \label{univresstr}
\frac{1}{\mu k}\sqrt{1+\mu^2k^2} \to
\frac{1}{\mu k}\biggl(\sqrt{1+\mu^2k^2} - \mu k \biggr),
\eeq
\[
\biggl(\mu k \sqrt{1+\mu^2k^2} - \mu^2 k^2 \biggr)\to
\biggl(\mu k \sqrt{1+\mu^2k^2} - \mu^2 k^2 - \frac{1}{2}\biggr).
\]

\noindent
In the first case we subtract 1, in the second case we subtract
$1/2$. In both cases we add the terms corresponding to these
constants to the $\mu$-independent terms in the integrands.  After
this restructuring (when needed) we write the integrals
\eq{sigma1tot} and \eq{sigma2tot} as sums of what we will call $\mu$- and
$c$-integrals

\beq \label{mucdefin}
\Delta\epsilon=\Delta\epsilon^{(\mu)}+ \Delta\epsilon^{(c)}.
\eeq

Let us consider calculation of $\mu$-integrals. The integral over
$k$ in \eq{sigma1tot} converges, the integrand does not require
any restructuring, and the $\mu$-integral has the form

\beq
\Delta \epsilon_{\Sigma 1}^{\,(\mu)}=-3\int_0^1 {dx} \int_0^x
{dy} \int_0^1 \frac{{dv}}{1-v^2}v^2\biggl(1-\frac{v^2}{3}\biggr)
h_1(x,y)
\eeq
\[
\times\int_0^{\infty}\frac{{dk^2}}{(k^2+a_1^2)^2} \biggl(\mu k
\sqrt{1+\mu^2k^2} - \mu^2 k^2 \biggr).
\]

\noindent
This integral is of order $\mu$, and therefore with our accuracy it does not give any contribution

\beq  \label{musigma1res}
\Delta \epsilon_{\Sigma 1}^{\,(\mu)} = 0.
\eeq

\noindent
Next we calculate the $\mu$-integral arising from the integral in \eq{sigma2tot}

\beq  \label{murestssigm2}
\Delta \epsilon_{\Sigma 2}^\mu= \int_0^1 {dx} \int_0^x {dy} \int_0^1
\frac{{dv}}{1-v^2}v^2\biggl(1-\frac{v^2}{3}\biggr) h_2(x,y)
\int_0^{\infty}{dk^2}\biggl\{\frac{1}{k^2+a_1^2}
\eeq
\[
\times\biggl[\frac{1}{\mu k}\biggl(\sqrt{1+\mu^2k^2}-\mu k \biggr)
- \frac{1}{2}\biggl(\mu k \sqrt{1+\mu^2k^2}- \mu^2 k^2 -\frac{1}{2}
\biggr)\biggl].
\]

\noindent
Unlike the case of $\Delta \epsilon_{\Sigma 1}^\mu$ this time the
naive integral over $k$ of the $\mu$-dependent terms in
\eq{sigma2tot} diverges at large $k$, and we restructured the
integrand according to \eq{univresstr}. Besides the recoil
contribution the integral in \eq{murestssigm2} contains also
nonrelativistic contribution $\Delta \epsilon_{\Sigma}^\mu(NR)$ that we
calculated separately in \eq{nonrelsigma}. This nonrelativistic contribution is generated by the leading $1/(\mu k)$ term in the expansion over small $\mu k$ of the expression in the square brackets in \eq{murestssigm2}. It coincides with the contribution generated by the first term in the square brackets in \eq{nonrelsigmares}. To avoid double counting we subtract this contribution from the integrand in \eq{murestssigm2}  by the substitution in the integrand

\beq  \label{munonrelsubtr}
\frac{1}{\mu k}~\sqrt{1+\mu^2k^2} \to
\frac{1}{\mu k}\biggl(\sqrt{1+\mu^2k^2} - 1 \biggr).
\eeq

\noindent
This substitution gives a universal recipe for subtraction of the
nonrecoil corrections in all $\mu$-integrals to be considered below,
and it is necessary to emphasize that it is needed only in the first
of two typical structures with square roots in \eq{univresstr} that
arise in the expressions for $\mu$-integrals. The leading term in the small $\mu k$ expansion of the second structure is nonsingular and does not generate nonrecoil contribution.

Finally, the second $\mu$-integral for the mass operator insertion in the electron line has the form

\beq
\Delta \epsilon_{\Sigma 2}^{\,(\mu)} - \Delta \epsilon_{\Sigma 2}^{\,(\mu)}(NR)
=\int_0^1 {dx} \int_0^x {dy}
\int_0^1 \frac{{dv}}{1-v^2}v^2\biggl(1-\frac{v^2}{3}\biggr) h_2(x,y)
\int_0^{\infty} \frac{{dk^2}}{k^2+a_1^2}
\eeq
\[
\times \biggl[~\frac{1}{\mu k}\biggl(\sqrt{1+\mu^2k^2} - \mu k - 1
\biggr) - \frac{1}{2}\biggl(\mu k \sqrt{1+\mu^2k^2} - \mu^2 k^2 -
\frac{1}{2} \biggr)\biggr].
\]

\noindent
Other integrals of this type will arise below in calculations of other contributions to HFS. To extract the first term in the expansion of integrals of this type over the small parameter $\mu$ we introduce an auxiliary parameter $\sigma$, such that $1\ll\sigma\ll1/\mu$ (see, e.g., \cite{berlifpit,yaf1988}). Then we separate the large and small integration momenta regions with the help of this parameter $\sigma$ and use different approximations in different regions. In the region of small integration momenta $0\leq k\leq \sigma$ we expand the integrand over $\mu k\ll 1$ and obtain

\begin{equation}
\label{smallsigma}
\Delta \epsilon_{\Sigma 2}^{\,(\mu \, <)} - \Delta E_{\Sigma 2}^{\,(\mu \, <)}(NR)
\simeq -\frac16 \ln^3{\sigma}
+ \frac{19}{24}\ln^2{\sigma} + \biggl(-\pi^2 +\frac{589}{72}\biggr)\ln{\sigma}
+ 1.3220.
\end{equation}

\noindent
In the region of large integration momenta $k\leq\sigma$ we expand the integrand over the small $1/k\ll1$ and obtain

\begin{equation}
\label{largesigma}
\Delta \epsilon_{\Sigma 2}^{\,(\mu \, >)}- \Delta \epsilon_{\Sigma 2}^{\,(\mu \, >)}(NR)
\simeq \frac16 \ln^3{(2\mu)}
+ \frac{1}{24}\ln^2{(2\mu)}+ \biggl(\frac{13\pi^2}{12} -
\frac{335}{36}~\biggr)\ln{(2\mu)}
\end{equation}
\[
+ \frac14\zeta{(3)} - \frac{215\pi^2}{144}+\frac{103}{8}
+\frac16\ln^3{\sigma}
- \frac{19}{24} \ln^2{\sigma} + \biggl(\pi^2 -
\frac{589}{72}\biggr)\ln{\sigma}.
\]

\noindent
In the intermediate region $k\sim\sigma$ both approximations $\mu k\ll 1$ and $1/k\ll1$ are valid simultaneously, and all dependence on the auxiliary parameter $\sigma$ cancels in the sum of the contributions in \eq{smallsigma} and \eq{largesigma}

\beq
\label{sigmamutotal}
\Delta \epsilon_{\Sigma 2}^{\,(\mu)} - \Delta \epsilon_{\Sigma 2}^{\,(\mu)}(NR)
\simeq \frac16\ln^3{(2\mu)}
+ \frac{1}{24}\ln^2{(2\mu)} + \biggl(\frac{13\pi^2}{12} -
\frac{335}{36}\biggr)\ln{(2\mu)}
\eeq
\[
+ \frac14 \zeta{(3)} - \frac{215\pi^2}{144}+\frac{103}{8}
+ 1.3220.
\]

Let us turn to calculation of the $c$-integrals as defined in \eq{mucdefin}, \eq{sigma1tot}, and \eq{sigma2tot}. We easily perform momentum integration in the integral for $\Delta \epsilon_{\Sigma 1}^{\,(c)}$

\beq  \label{cintsigma1unsub}
\Delta \epsilon_{\Sigma 1}^{\,(c)}
=\frac{3}{4}\int_0^1 {dx} \int_0^x
{dy} \int_0^1 \frac{{dv}}{1-v^2}v^2\biggl(1-\frac{v^2}{3}\biggr)
h_1(x,y)\int_0^{\infty}\frac{dk^2}{k^2}
\eeq
\[
\times\biggl[\frac{1}{k^2+a_1^2}\sqrt{(k^2+a_1^2)^2+4k^2}
-1\biggr]
\]
\[
=3\int_0^1 {dx} \int_0^x {dy}
\int_0^1 \frac{{dv}}{1-v^2}v^2\biggl(1-\frac{v^2}{3}\biggr) h_1(x,y)
\biggl[\frac{1}{a_1}\tan^{-1}\frac{1}{a_1} -
\frac12\ln{\frac{1+a_1^2}{  a_1^2}}\biggr].
\]

\noindent
As in the case of the $\mu$-integrals above we want to avoid double counting, and subtract the respective nonrelativistic contribution already accounted for in \eq{nonrelsigmares}. This nonrelativistic contribution has the form

\beq
\Delta \epsilon_{\Sigma 1}^{\,(c)}(NR)=3\int_0^1 {dx} \int_0^x {dy}
\int_0^1 \frac{{dv}}{1-v^2}~v^2\biggl(1-\frac{v^2}{3}\biggr)
\frac{h_1(x,y)}{a_1}.
\eeq

\noindent
Now we see that due to the identity $\tan^{-1}(1/a_1)=\pi/2-\tan^{-1}{a_1}$ subtraction of nonrelativistic contribution from the integral in \eq{cintsigma1unsub} reduces to substitution

\beq \label{cintmassshsusbt}
\tan^{-1}\left(\frac{1}{a_1}\right)\to -\tan^{-1}{a_1}.
\eeq

\noindent
This is a universal rule for subtraction of nonrelativistic contributions in all $c$-integrals considered below (compare \cite{jetp92}). Finally, the first $c$-integral is

\beq \label{cintsigma1res}
\Delta \epsilon_{\Sigma 1}^{\,(c)}-\Delta \epsilon_{\Sigma 1}^{\,(c)}(NR)=
\int_0^1 {dx} \int_0^x {dy}
\int_0^1 \frac{{dv}}{1-v^2}v^2\biggl(1-\frac{v^2}{3}\biggr) h_1(x,y)
\eeq
\[
\times\biggl[-\frac{3}{a_1}\tan^{-1}{a_1}-
\frac32\ln{\frac{1+a_1^2}{a_1^2}} \biggr]
= -2.6215.
\]

Next we turn to the second $c$-integral defined in \eq{mucdefin} and \eq{sigma2tot}, and start with the momentum integration (compare \cite{yaf1988})

\beq
\Delta \epsilon_{\Sigma 2}^{\,(c)}
= \int_0^1 {dx} \int_0^x {dy} \int_0^1
\frac{{dv}}{1-v^2}v^2\biggl(1-\frac{v^2}{3}\biggr) h_2(x,y)
\int_0^{\infty}{dk^2}
\eeq
\[
\times\biggl[\frac{(k^2+a_1^2)-\sqrt{(k^2+a_1^2)^2+4k^2}}{(k^2+a_1^2)^2}
+\frac{1}{8k^2}\biggl(\sqrt{(k^2+a_1^2)^2+4k^2}-(k^2+a_1^2)
-\frac{2k^2}{k^2+a_1^2}
\biggr)\biggl]
\]
\[
= \int_0^1 {dx} \int_0^x {dy} \int_0^1
\frac{{dv}}{1-v^2}v^2\biggl(1-\frac{v^2}{3}\biggr) h_2(x,y)
\biggl[-\frac{2}{a_1}\tan^{-1}\frac{1}{a_1}
+\frac{3}{4}\ln{\frac{1+a_1^2}{a_1^2}}
+ \frac14-\frac{a_1^2}{4}\ln{\frac{1+a_1^2}{a_1^2}}\biggr].
\]

\noindent
Subtraction of the nonrecoil contribution is needed to avoid double counting, and it is done with the help of the universal rule in \eq{cintmassshsusbt}. Then we obtain

\beq \label{cintsigma2res}
\Delta \epsilon_{\Sigma 2}^{\,(c)} - \Delta \epsilon_{\Sigma 2}^{\,(c)}(NR)
= \int_0^1 {dx} \int_0^x {dy}
\int_0^1 \frac{{dv}}{1-v^2}v^2\biggl(1-\frac{v^2}{3}\biggr)h_2(x,y)
\biggl[\frac{2}{a_1}\tan^{-1}{a_1}
\eeq
\[
+\frac{3}{4}\ln{\frac{1+a_1^2}{a_1^2}}+\frac14
-\frac{a_1^2}{4}\ln{\frac{1+a_1^2}{a_1^2}} \biggr]
= 0.4370.
\]

Next we collect all contributions in \eq{nonrelsigmares}, \eq{musigma1res}, \eq{sigmamutotal}, \eq{cintsigma1res}, and \eq{cintsigma2res}, and obtain the total result for the contribution of the diagram with the self-energy insertion in Fig.~\ref{elline}

\beq \label{totsigmeel}
\Delta\epsilon_\Sigma
=\Delta \epsilon_{\Sigma}(NR)
+(\Delta\epsilon_{\Sigma 1}^{\,(\mu)}- \Delta \epsilon_{\Sigma 1}^{\,(\mu)}(NR))
+(\Delta \epsilon_{\Sigma 2}^{\,(\mu)} - \Delta \epsilon_{\Sigma 2}^{\,(\mu)}(NR))
\eeq
\[
+(\Delta \epsilon_{\Sigma 1}^{\,(c)} - \Delta \epsilon_{\Sigma 1}^{\,(c)}(NR))
+(\Delta \epsilon_{\Sigma 2}^{\,(c)} - \Delta \epsilon_{\Sigma 2}^{\,(c)}(NR))
=\frac{0.4462}{\mu}
\]
\[
+\frac16\ln^3{(2\mu)}
+\frac{1}{24} \ln^2{(2\mu)} +\biggl(\frac{13\pi^2}{12}-
\frac{335}{36}\biggr)\ln{(2\mu)} - 0.1856.
\]

\subsection{Calculation of the Spanning Photon Contribution}

Contribution of the spanning photon diagrams in Fig.~\ref{elline} is obtained from the contribution of the skeleton diagrams in Fig.~\ref{twoph} by insertion of the radiative photon with the polarization bubble, and is described by the expression (compare \cite{yaf1988})

\beq
\Delta \epsilon_{\,\Xi} =-\frac{i}{16\pi^2\mu^2}
\int_0^1 {dx} \int_0^x {dy}(x-y)
\int_0^1 \frac{{dv}}{1-v^2}v^2\biggl(1-\frac{v^2}{3}\biggr)~
\int \frac{{d^4k}}{k^4}
\eeq
\[
\times\biggl(\frac{1}{k^2+\mu^{-1}k_0+i0}
+\frac{1}{k^2-\mu^{-1}k_0+i0}\biggl)~\Biggl\{2(3k_0^2 - 2{\bm k}^2)
\biggl[ \frac{1-3y}{\Delta}
\]
\[
+\frac{-k^2y^2(1-y)+2bk_0y^2(1-y)-2+x(2-x)(1-y)}{\Delta^2}
\biggr]
-6bk_0 \biggl[ \frac{3(1-y)}{\Delta}
\]
\[
+\frac{k^2y(1-y)(2-y)-2bk_0y(1-y)^2
+x(2-x)(1-y)}{\Delta^2}\biggr]\Biggr\},
\]

\noindent
where

\beq \label{spanaux}
\Delta(x,y) = y(1-y)(-k^2 + 2bk_0 + a^2 -i0),
\eeq
\[
a^2(x, y) = \frac{x^2+\lambda^2 (1-x)}{y(1-y)},\quad
b(x, y) = \frac{1-x}{1-y}.
\]

\noindent
We simplify this expression using the identities

\beq
-k^2y^2(1-y)+2bk_0y^2(1-y)+x(2-x)(1-y)
=y\Delta+2x(1-y)-x^2-\lambda^2(1-x)y,
\eeq
\[
k^2y(1-y)(2-y)-2bk_0y(1-y)^2+x(2-x)(1-y)
\]
\[
=-(2-y) \Delta + 2bk_0y(1-y)+2x(1-y)+x^2+\lambda^2(1-x)(2-y),
\]

\noindent
and obtain

\beq \label{spanrans}
\Delta \epsilon_{\,\Xi} =-\frac{i}{16\pi^2\mu^2}
\int_0^1 {dx} \int_0^x {dy}(x-y)
\int_0^1 \frac{{dv}}{1-v^2}v^2\biggl(1-\frac{v^2}{3}\biggr)
\int \frac{{d^4k}}{k^4}\biggl(\frac{1}{k^2+\mu^{-1}k_0+i0}
\eeq
\[
+\frac{1}{k^2-\mu^{-1}k_0+i0}\biggl)\Biggl\{2(3k_0^2 - 2{\bm k}^2)
\biggl[ \frac{1-2y}{\Delta}+\frac{-2+2x(1-y)-x^2}{\Delta^2}
+\frac{-\lambda^2 (1-x)y}{\Delta^2}\biggr]
\]
\[
-6bk_0 \biggl[ \frac{1-2y}{\Delta}+\frac{2x(1-y)+x^2}{\Delta^2}
+\frac{\lambda^2 (1-x)(2-y)}{\Delta^2}+\frac{2bk_0 y(1-y)}{\Delta^2}
\biggr]\Biggr\}\equiv\Delta \epsilon_{\,\Xi1}+\Delta \epsilon_{\,\Xi2},
\]

\noindent
where the contributions $\Delta \epsilon_{\,\Xi1}$ and $\Delta \epsilon_{\,\Xi2}$ correspond to the expressions in the first and second square brackets on the right hand side in \eq{spanrans}.

\begin{table}
\caption{Spanning Photon Contributions}
\begin{ruledtabular}
\begin{tabular}{l|c|r|}
&$\Xi1$ & $\Xi2$ \\
\hline

$\Delta \epsilon(NR)$ &$\frac{0.41386}{\mu}-0.1186$ & $0.7213$\\

$\Delta \epsilon^{(\mu)}-\Delta \epsilon(NR)$ &$\frac{1}{6}\ln^3(2\mu)
-\frac{5}{24}\ln^2(2\mu) +\biggl(\frac{\pi^2}{12}+
\frac{13}{36}\biggr)\ln(2\mu)$
$+\frac{1}{4}\zeta(3)- \frac{5\pi^2}{144}-\frac{9}{8}
+0.5259$
&0\\

$\Delta \epsilon^{(c)}-\Delta \epsilon(NR)$ & $0.1201$ & $-0.7130$ \\


\end{tabular}
\end{ruledtabular}
\end{table}

For calculation of the of the integrals in \eq{spanrans} we use the same tricks as in the case of the mass operator contribution. We skip the details of calculations and collect intermediate results in Table 1. Finally we obtain the contribution of the diagram with the spanning photon insertion in Fig.~\ref{elline} in the form

\beq  \label{spaneltotal}
\Delta \epsilon _\Xi=
\frac{0.4139}{\mu}+\frac{1}{6}\ln^3(2 \mu) - \frac{5}{24}\ln^2(2 \mu) + \biggl(\frac{\pi^2}{12} + \frac{13}{36}\biggr)\ln(2 \mu)-0.6314.
\eeq

\subsection{Calculation of the Vertex Contribution}

Contribution of the diagram with the vertex insertion in Fig.~\ref{elline} is obtained from the contribution of the skeleton diagrams in Fig.~\ref{twoph} by substitution of the vertex function instead of one of the skeleton vertices. We have derived a convenient expression for the one-loop vertex function with a massive photon

\beq \label{vert2004}
\Lambda_{\mu} =\frac{\alpha}{2\pi}\int_0^1 {dx} \int_0^x
{\frac{dy}{\Delta}}\gamma_{\mu}\Biggl\{(k^2-2k_0)
\Bigl[(x-y)(1-2y)+ y(1-y)\Bigr]
+ 2 \biggl(1-x-\frac{x^2}{2}\biggr)\frac{\Delta -\Delta_0}{\Delta_0}
\eeq
\[
+ ({\crossed p}-m)\,x(1-x)\frac{\Delta -\Delta_0}{\Delta_0}
+2k_0 \Bigl[~1-x + (x-y)^2\Bigr]
-({\crossed p}-{\crossed k}-m)(1-x)\Biggr\}
=\sum_{i=1}^{i=5}\Lambda_{\mu}^{(i)},
\]

\noindent
where $\Delta_0(x) =x^2+\lambda^2 (1-x)$, and the terms $\Lambda_{\mu}^{(i)}$ correspond to the five terms in the braces in \eq{vert2004}.

This is essentially the same expression as the one in \cite{yaf1988}. But unlike the respective expression in \cite{yaf1988}, where the photon mass merely served as a regularization parameter and was preserved only when necessary, here we have restored full dependence on the finite photon mass $\lambda$ that effectively describes polarization operator insertion. One can show that the gauge invariant anomalous magnetic moment does not generate radiative-recoil corrections (see, e.g., \cite{jetp92,eksann1}). Therefore, the anomalous magnetic moment and some other terms that do not contribute to HFS are omitted in \eq{vert2004}.

\begingroup
\squeezetable
\begin{table}
\caption{Vertex Contributions}
\begin{ruledtabular}
\begin{tabular}{l|l|l|l|}
&$\Delta \epsilon(NR)$  & $\Delta \epsilon^{(\mu)}-\Delta \epsilon(NR)$
& $\Delta \epsilon^{(c)}-\Delta \epsilon(NR)$  \\
\hline

$\Lambda1$ &$-\frac{1.1356}{\mu} + 0.3385$
& $-\frac{1}{6}\ln^3(2\mu)
+ \frac{11}{24}\ln^2(2\mu)+\biggl(-\frac{5\pi^2}{12} +
\frac{19}{8}\biggr)\ln{(2\mu)} + 3.2201$
& $-0.3315$\\

$\Lambda2$ &
$-\frac{0.0507}{\mu} +  0.1007$
& $\biggl(\frac{2\pi^2}{3}-\frac{469}{72}\biggr)\ln\frac{M}{m}
+ 0.0371$&  $-0.0604$\\

$\Lambda3$
& $-\frac{0.0104}{\mu} + 0.0160$
& $-3 \biggl( \frac{\pi^2}{3} -\frac{119}{36}\biggr)$
& $-0.0106$\\

$\Lambda4$& $3.0460$& $0$& $-2.4323$\\

$\Lambda5$& $-1.1758$& $0$& $1.1540$\\

\end{tabular}
\end{ruledtabular}
\end{table}
\endgroup

We plugged in the vertex in \eq{vert2004} in the skeleton expression for the contribution to HFS, and obtained five integrals corresponding to the five terms on the right hand side in \eq{vert2004}. Calculation of these integrals goes along the same lines as in the case of the mass operator, discussed in detail above. We collected all intermediate results in Table II. Total contribution of the diagram with the vertex insertion in Fig.~\ref{elline} turns out to be

\beq \label{totvertel}
\Delta \epsilon_{\Lambda}=-\frac{1.1968}{\mu}
-\frac{1}{6}\ln^3(2\mu)
+\frac{11}{24}\ln^2(2\mu) + \biggl(-\frac{13\pi^2}{12} +
\frac{80}{9}\biggr)\ln(2\mu) +3.9489.
\eeq

Now we collect all contributions in \eq{totsigmeel}, \eq{spaneltotal}, and \eq{totvertel}  generated by the diagrams with electron polarization insertions in Fig.~\ref{elline}, and obtain

\beq \label{fianlelctrless}
\Delta \epsilon^{(e)}=\Delta \epsilon_{\Sigma}+2\Delta \epsilon_{\Lambda}+\Delta \epsilon_{\Xi}
=-\frac{1.5335}{\mu}
+ \frac{3}{4}\ln^2(2\mu) + \biggl(-\pi^2 +
\frac{53}{6}\biggr)\ln(2\mu) + 7.0807.
\eeq

\noindent
The first term on the right hand side is the well known nonrecoil contribution to HFS \cite{eks2}, the second term is the leading logarithm squared contribution obtained in \cite{eks89}, and the single-logarithmic and constant terms are the subject of this work.

\section{Muon Polarization Operator}

Insertion of the muon polarization operator in the radiative photons in the diagrams in Fig.~\ref{muline} lifts characteristic integration momenta to the scale of the muon mass. Hence, these  diagrams do not generate nonrecoil contributions to HFS, all of which originate from the region of nonrelativistic muon momenta. Moreover, due to high characteristic momenta these diagrams even do not generate logarithmic in the mass ratio recoil contributions that originate from the wide integration region between the electron and muon mass. As a result leading recoil contributions of the diagrams in Fig.~\ref{muline} are pure numbers, and their calculation is significantly simpler than in the case of the electron polarization insertions in Fig.~\ref{elline}.

Like in the previous section, insertion of the muon polarization operator in the diagrams in Fig.~\ref{muline} is accounted for by introduction of a photon mass, followed by an additional integration over the velocity with the weight from \eq{velintweight}. For muon polarization the effective photon mass in the integrals is large, in dimensionful units $\lambda^2=4M^2/(1-v^2)$, and it determines characteristic momenta in all integrals.  We can obtain the expressions for energy shift due to muon polarization from the formulae for the respective electron polarization contributions above by rescaling the dimensionless integration momenta $k\to k/\mu$ (recall that $\mu=m/(2M)$). In addition we should adjust the expression for the photon mass, in terms of the rescaled integration momenta measured in units of $2M$ it is $\lambda^2=1/(\mu^2(1-v^2))$. After these substitutions the expressions for the dimensionless contributions to HFS that are due to electron polarization go into expressions for the contributions due to muon polarization.

\paragraph{Mass operator contribution.} To obtain an explicit expression for the diagrams with the mass operator insertions in Fig.~\ref{muline} we rescale the integration momentum $k\to k/\mu$ in \eq{sigma1tot} and \eq{sigma2tot}, and redefine the photon mass squared as $\lambda^2=1/(\mu^2(1-v^2))$. The auxiliary functions used in \eq{sigma1tot} and \eq{sigma2tot} are defined in \eq{auxsigmele}. After rescaling they  simplify

\beq
h_2(x,y) \to \frac{1-x}{y},\qquad
a_1(x,y)\to\frac{1}{y\mu^2(1-v^2)}.
\eeq

\noindent
All dependence on $\mu$ becomes explicit after these manipulations. It turns out that $\Delta\epsilon_{\Sigma1}$ vanishes together with $\mu$. The total leading recoil contribution generated by the diagrams with mass operator insertions in Fig.~\ref{muline} coincides with $\Delta\epsilon_{\Sigma2}$. Its calculation is straightforward, and we obtain

\beq \label{sigmsmuon}
\Delta \epsilon_{\Sigma }=\int_0^1 {dx} \int_0^x {dy}
\int_0^1 {dv}~v^2\biggl(1-\frac{v^2}{3}\biggr)
\int_0^{\infty}{dk^2}~\frac{1-x}{k^2y(1-v^2)+1}
\eeq
\[
\times\biggl[\frac{1}{k}\biggl(\sqrt{1+k^2} - k  \biggr)
- \frac{1}{2}\biggl(k \sqrt{1+k^2} - k^2
- \frac{1}{2} \biggr)\biggr]=0.1329.
\]

\paragraph{Spanning photon contribution.} We obtain an expression for the spanning photon contribution with the muon polarization insertion in Fig.~\ref{muline} by rescaling the integration momentum  and the photon mass in \eq{spanrans}. Under these transformations the auxiliary function $\Delta(x,y)$ in \eq{spanaux} simplifies

\beq
\Delta(x,y) \to \frac{y(1-y)}{\mu^2} \biggl[-k^2 +\frac{\mu^2 \lambda^2\,(1-x)}{y(1-y)}\biggr],
\eeq

\noindent
After rescaling only the first and third terms in the first square bracket on the right hand side in \eq{spanrans} produce nonvanishing with $\mu$ contributions, and we obtain
\beq  \label{spanmuon}
\Delta \epsilon_{\Xi}
=\int_0^1 {dx} \int_0^x {dy}
\int_0^1 {dv}v^2\biggl(1-\frac{v^2}{3}\biggr)
\int_0^{\infty} {dk^2} (x-y)
\eeq
\[
\times\biggl[\frac{1-2y}{k^2 y(1-y)(1-v^2)+1-x}
-\frac{(1-x)y}{[k^2 y(1-y)(1-v^2)+1-x]^2}\biggr]
\]
\[
\times \biggl[\frac{1}{k}\biggl(\sqrt{1+k^2} - k  \biggr)
- \frac{1}{2}\biggl(k \sqrt{1+k^2} - k^2
- \frac{1}{2} \biggr)\biggr]= 0.3105.
\]

\paragraph{Vertex contribution.}
Rescaling the integration momentum and the photon mass in the expressions corresponding to the five terms $\Lambda^{(i)}_\mu$ we obtain an explicit expression for the vertex diagram contribution in Fig.~\ref{muline} to HFS. It is easy to see that after rescaling all terms in the expression for the vertex function in \eq{vert2004} are suppressed by at least one power of $\mu$ in comparison with the first term. This means that only the first term generates the leading recoil correction in the case of muon polarization insertion. Explicitly, the leading recoil correction generated by the vertex insertions in Fig.~\ref{muline} has the form

\beq \label{vertmuon}
\Delta \epsilon_{\Lambda }
=- \int_0^1 {dx} \int_0^x {dy}
\int_0^1 {dv}v^2\biggl(1-\frac{v^2}{3}\biggr)
\int_0^{\infty} {dk^2}\frac{(x-y)(1-2y)+y(1-y)}{k^2 y(1-y)(1-v^2)+1-x}
\eeq
\[
\times \biggl[~\frac{1}{k}\biggl(\sqrt{1+k^2} - k  \biggr)
- \frac{1}{2}\biggl(k \sqrt{1+k^2} - k^2
- \frac{1}{2} \biggr)\biggr] =-0.~8738.
\]

Collecting all leading radiative-recoil corrections in \eq{sigmsmuon}, \eq{spanmuon}, and \eq{vertmuon} corresponding to the diagrams with muon polarization insertions in Fig.~\ref{muline}, we obtain

\beq \label{fianlmuontrless}
\Delta\epsilon^{(\mu)}=\Delta\epsilon_{\Sigma} + 2 \Delta \epsilon_{\Lambda}
+\Delta \epsilon_{\Xi}=-1.3042.
\eeq

\section{Conclusions}

Restoring the overall dimensional factor in \eq{fianlelctrless} and throwing away nonrecoil and logarithm squared terms known earlier, we  obtain single-logarithmic and nonlogarithmic contributions to HFS generated by the diagrams with one-loop electron polarization insertions in
Fig.~\ref{elline}

\begin{equation}
\Delta E^{(e)}=
\Biggl[\biggl(\pi^2 -\frac{53}{6}~\biggr)\ln{\frac{M}{m}} +
7.0807\Biggr]\frac{\alpha^2(Z\alpha)}{\pi^3}\frac{m}{M}E_F .
\end{equation}

\noindent
The radiative-recoil contribution generated by the diagrams with one-loop muon polarization insertions in Fig.~\ref{muline} is nonlogarithmic. We obtain it restoring the overall dimensional factor in \eq{fianlmuontrless}

\begin{equation}
\Delta E^{(\mu)}
=-1.3042~\frac{\alpha(Z^2\alpha)(Z\alpha)}{\pi^3}\frac{m}{M}E_F.
\end{equation}

\noindent
The total contribution to HFS obtained above can be written as ($Z=1$ in muonium)

\beq \label{totress}
\Delta E=\Delta E^{(e)}+\Delta E^{(\mu)}
=\Biggl[\biggl(\pi^2 -\frac{53}{6}~\biggr)\ln{\frac{M}{m}} +
5.7765\Biggr]\frac{\alpha^3}{\pi^3}\frac{m}{M}E_F.
\eeq

Currently the theoretical accuracy of HFS in muonium is about 70 Hz. A realistic goal is to reduce this uncertainty below 10 Hz (see a more detailed discussion in \cite{egsbook,preprts}). The result in \eq{totress} together with other three-loop radiative-recoil results in \cite{egs01,egs03,egs04} makes this goal closer.

\begin{acknowledgments}
This work was supported by the NSF grant PHY-0757928. V.A.S. was also
supported in part by the RFBR grants 06-02-16156 and 08-02-13516, and by the DFG grant GZ  436 RUS 113/769/0-3.
\end{acknowledgments}

\end{document}